\documentclass{article}
\usepackage{graphicx}
\usepackage{epstopdf}
\usepackage{amssymb,amsmath}
\usepackage{amsthm}
\usepackage{url}
 
\begin{document}
 \pagestyle{empty}
\title{\bf Gap-Measure Tests with Applications to\\
       Data Integrity Verification}
\author{{\bf Truc Le} and {\bf Jeffrey Uhlmann}\\Department of Computer Science\\University of Missouri - Columbia}
\date{}
 \maketitle
\begin{abstract}
In this paper we propose and examine gap statistics for assessing uniform distribution hypotheses. We provide examples relevant to data integrity testing for which max-gap statistics provide greater sensitivity than chi-square ($\chi^2$), thus allowing the new test to be used in place of or as a complement to $\chi^2$ testing for purposes of distinguishing a larger class of deviations from uniformity. We establish that the proposed max-gap test has the same sequential and parallel computational complexity as $\chi^2$ and thus is applicable for Big Data analytics and integrity verification.
\end{abstract}

\section{Introduction}
\label{sec:intro}
Distribution testing is a fundamental statistical problem that arises in a wide range of practical applications. At its core the problem is to assess whether a dataset that is assumed to comprise samples from a known probability distribution is in fact consistent with that assumption. For example, if the end state of a computer simulation of a physical system is a set of points with an expected physics-prescribed distribution, then any detected deviation from that expected distribution could undermine confidence in the results obtained and possibly in the integrity of the simulation system itself.
 
Data integrity verification is a related application for distribution testing in which the objective is to detect evidence of tampering, e.g., human-altered data. For example, many sources of numerical data produce numbers with first digits conforming to the Benford-Newcomb first-digit distribution\footnote{This phenomenon is often referred to as ``Benford's Law''.\newline\newline
Manuscript received April 19, 2005; revised September 17, 2014. Published: {\em Statistics Research Letters},
vol.\ 4, pp.\ 11-17, 2015.} \cite{Nigrini2012,Winter2012}, while digits other than the first and last are uniformly sampled from ${0,...,9}$ \cite{Dlugosz2009}. Digits in human-created numbers, by contrast, tend to exhibit high regularity with all elements of ${0,...,9}$ represented with nearly equal cardinality. Statistically identified deviations of this kind have been used to uncover acts of scientific misconduct and accounting fraud \cite{Pirracchio2013,Bolton2002,Kingston2014,Nigrini2011,Diekmann2011,Leemann2014}, but there is an increasing need for higher-sensitivity tests.
 
There is of course no way to make an unequivocal binary assessment of whether a dataset of samples conforms to a given distribution assumption, but it is possible to devise statistical tests which can assign a rigorous likelihood estimate to the hypothesis that the dataset does (or does not) represent samples from the assumed distribution. In this paper we briefly review the most widely-used method for distribution testing, the chi-square ($\chi^2$) test, and then develop alternative tests based on the statistics of gap-widths between data items of consecutive rank. Our principal contribution is a max-gap test which is shown to provide superior sensitivity to regularity deviations from a uniform distribution that are relevant to data integrity testing \cite{Fujii2012,Al-Marzouki2005,Simonsohn2013}. We show that this test can be evaluated with the same optimal computational complexity (serial and parallel) as the conventional $\chi^2$ test and is therefore suitable for extremely large-scale datasets.
 
\section{Chi-square Test}
The $\chi^2$ test is a statistical measure that can be applied to a discrete dataset to assess the hypothesis that its elements were sampled from a particular distribution. More specifically, it is a histogram-based method to measure the goodness-of-fit between the observed frequency distribution and the expected (theoretical) frequency distribution. The general procedure of the test includes the following steps:
\begin{enumerate}
\item Calculate the chi-square statistic, $\chi^2$, which is a normalized sum of squared differences (deviations) between observed and expected frequencies.
\item Determine the degrees of freedom, $df$, of that statistic, which is essentially the number of frequencies reduced by the number of parameters of the fitted distribution.
\item Compare $\chi^2$ to the critical value for the chi-square distribution with $df$ degrees of freedom.
\end{enumerate}
 
\begin{figure}
\centering
\includegraphics[scale = 0.25]{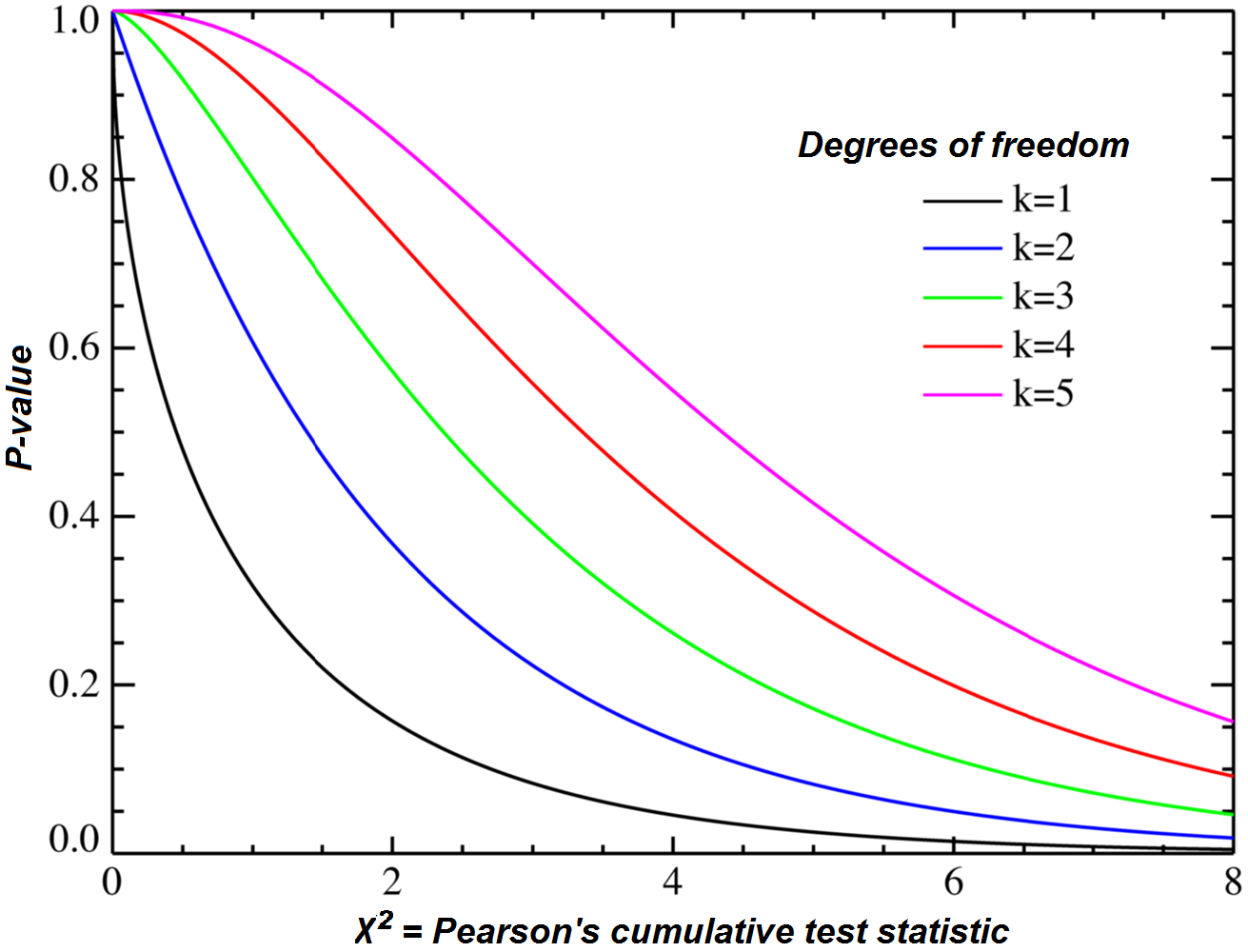}
\caption{Complement of the cumulative distribution function of the $\chi^2$ distribution, showing $\chi^2$ on the $x$-axis and $p$-value on the $y$-axis \cite{X2ICDF}.}
\label{fig:chi-square-complement-cdf}
\end{figure}
 
An example of the complement of the cumulative distribution function of the $\chi^2$ distribution is shown in Fig. \ref{fig:chi-square-complement-cdf} with different degrees-of-freedom values. For uniformity testing the procedure can be expressed as follows:
\begin{enumerate}
\item Given $N$ observations, construct an $N$-bin histogram. Let $b_i$ be the bin count for the $i^\text{th}$ bin ($i = 1, \dotsc, N$), which is the observed frequency distribution. As we are testing for uniformity, the expected frequency distribution $e_i = 1, \forall i = 1, \dotsc, N$.
\item Compute the chi-square test statistic:
\begin{equation}
\chi^2 = \sum\limits_{i = 1}^N \frac{(b_i - e_i)^2}{e_i} = \sum\limits_{i = 1}^N (b_i - 1)^2
\end{equation}
\item The number of degrees of freedom, {\em df}, is $N - 1$ for this case because the counts for $N - 1$ bins uniquely determine the count for the remaining bin.
\item Compute the complement of the cumulative distribution function of the $\chi^2$ distribution with $\chi^2$ and $df$ obtained from the previous steps. Compare this value with the significance level $\alpha$ for the test result.
\end{enumerate}
 
Despite being the de~facto standard for assessing dataset consistency with respect to a given distribution assumption, the $\chi^2$ test is not optimally sensitive to the types of deviation from uniformity that arise in many data integrity applications. One example involves narrow-band missing data resulting from a corrupted sensor or measurement process. Another example involves data that is generated from a non-random process and exhibits a higher degree of data regularity than is expected for a uniform distribution \cite{Pitt2013,Hill2014}. Datasets of the latter kind are typical of artificial and human-generated data, e.g., as in a forged dataset that has been tailored to include deviations that qualitatively resemble (to humans) uniform random deviates. In the following section we demonstrate the advantage of the proposed max-gap test over $\chi^2$ for narrow-band and high-regularity deviations from uniformity.
 
\section{Max-gap Test}
The maximum gap, or max-gap, for a dataset of real values is defined as the maximum difference between elements of consecutive rank, which can be determined from a sorted ordering of the dataset. The distribution of spacings between consecutive-rank items in a dataset has been examined in the literature \cite{Darling1953,Pyke1965,Pyke1972,Holst1980}, and we summarize here some of the results relevant to gap analysis. Assume we are given $N - 1$ observations on the open unit interval $(0, 1)$ which divide the interval into $N$ intervals whose lengths in ascending order are denoted by $S_{(1)} < S_{(2)} < \dotsb < S_{(N)}$. For uniformity testing we are interested in $S_{(N)}$, as it is the \emph{max-gap} of the observations. The exact distribution of $S_{(N)}$ is \cite{Holst1980}:
\begin{equation}
P(S_{(N)} \leq x) = \sum\limits_{\nu = 0}^N (-1)^\nu \binom{n}{\nu} (1 - nx)_{+}^{N - 1}
\label{eqn:exact-max-gap-distribution}
\end{equation}
where $a_+ = max(a, 0)$.\\
 
From the p-value of the max-gap $S_{(N)}$, denoted by $p$, we can perform a max-gap test for uniformity by checking the condition $p \geq \alpha$ for the \emph{one-sided} test, or $1 - \frac{\alpha}{2} \geq p \geq \frac{\alpha}{2}$ for the \emph{two-sided} test, where $\alpha$ is the significance level. When $N$ is large we may replace computation of the exact cumulative distribution of the max-gap in Eqn.~(\ref{eqn:exact-max-gap-distribution}) with the following asymptotic result \cite{Holst1980}:
\begin{equation}
P \left( S_{(N)} \leq x \right) \overset{N \rightarrow \infty}{=} e^{-e^{\ln N - Nx}},
\label{eqn:max-gap-distribution}
\end{equation}
where the expected value of $S_{(N)}$ is
\begin{equation}
E \left( S_{(N)} \right) \overset{N \rightarrow \infty}{=} \frac{\gamma + \ln N}{N}
\label{eqn:expected-max-gap}
\end{equation}
where $\gamma$ is Euler's constant.\\
 
An efficient max-gap test for uniformity can then be formalized as follows: Given $N - 1$ observations $x_i$, and a significance level $\alpha$, compute the max-gap $S_{(N)}$ of $\{0, 1\} \cup \{x_i\}$. Next, the p-value of the statistics is calculated as:
\begin{equation}
p = 1 - e^{-e^{\ln N - N S_{(N)}}}
\label{eqn:max-gap-p-value}
\end{equation}
 
If the p-value satisfies $p \geq \alpha$ for the one-sided test, or $1 - \frac{\alpha}{2} \geq p \geq \frac{\alpha}{2}$ for the two-sided test, the observations are deemed to pass the test. Otherwise the set of observations is assessed to be inconsistent with a uniform-sampling hypothesis and fails the test.
 
In the next section, we present results of experiments comparing the relative sensitivities of the $\chi^2$ test and the max-gap test for, e.g., indentifying anomalous regularity in a presumed-uniform distribution.
 
\section{Experiments}
\begin{figure}
\centering
\includegraphics[scale = 0.6]{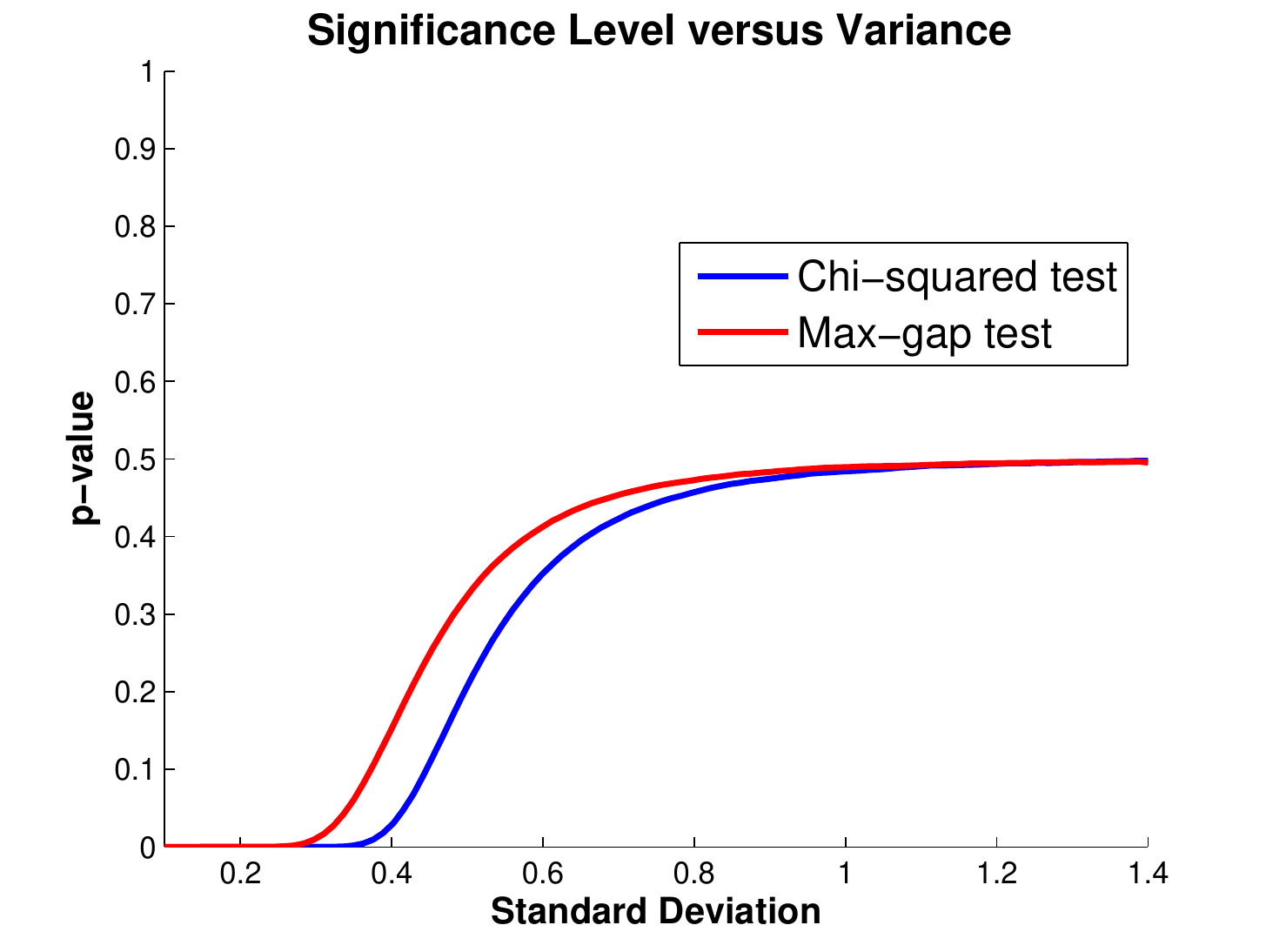}
\caption{The p-values of the $\chi^2$ test and the max-gap test of a normal distribution sampled within a fixed interval. When the standard deviation ($\sigma$) is small, both tests easily identify the data's non-uniformity. As $\sigma$ increases, the data distribution approaches uniformity within the sample interval and hence the p-values converge to $0.5$. This is an example in which the $\chi^2$ test should and does provide inherently greater sensitivity than the max-gap test.}
\label{fig:variance}
\end{figure}
 
\begin{figure}
\centering
\includegraphics[scale = 0.6]{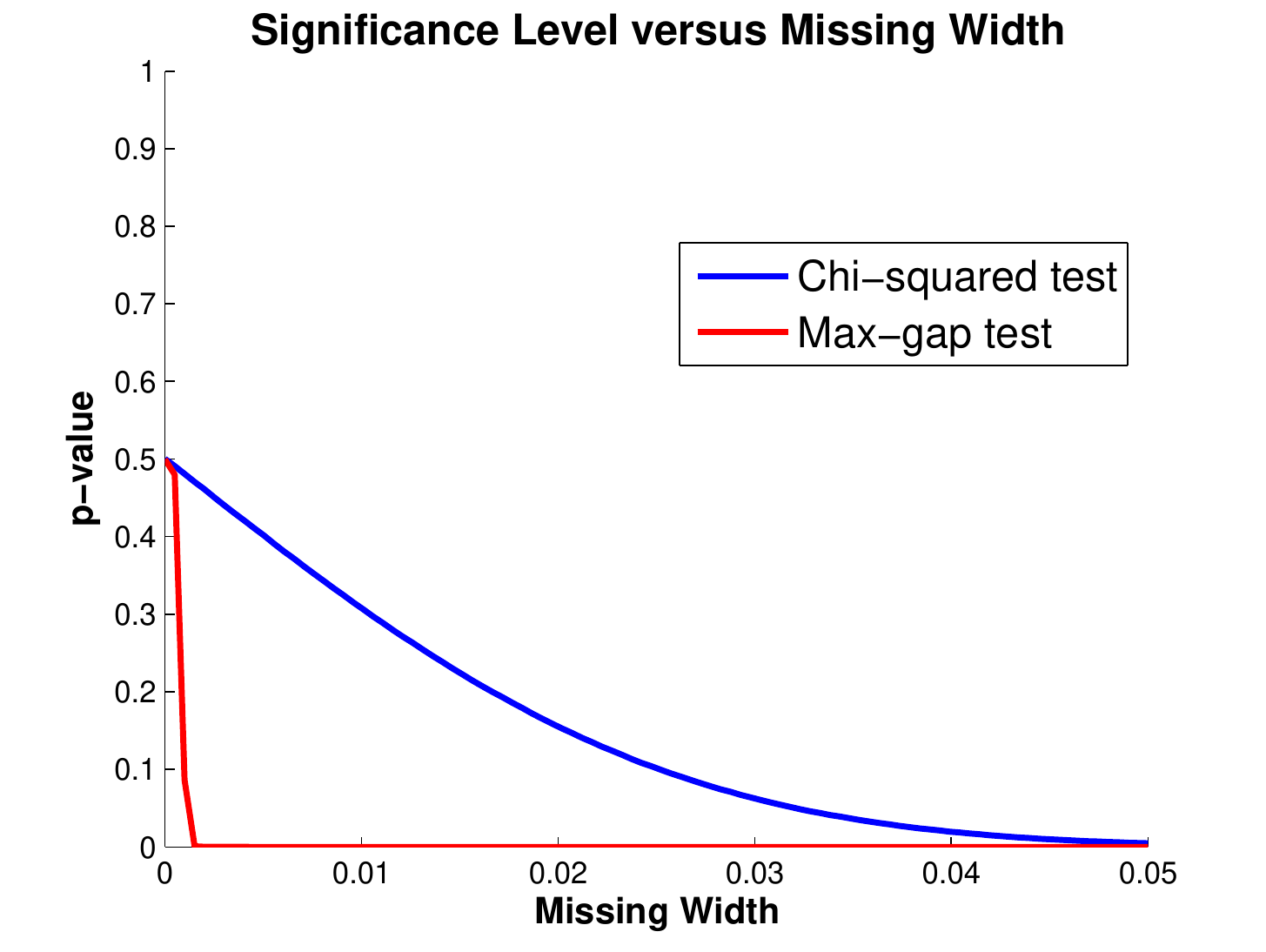}
\caption{p-value of the $\chi^2$ test and the max-gap test for narrow-band missing data. In this case the max-gap test provides inherently greater sensitivity.}
\label{fig:missing-values}
\end{figure}
 
\begin{figure}
\centering
\includegraphics[scale = 0.6]{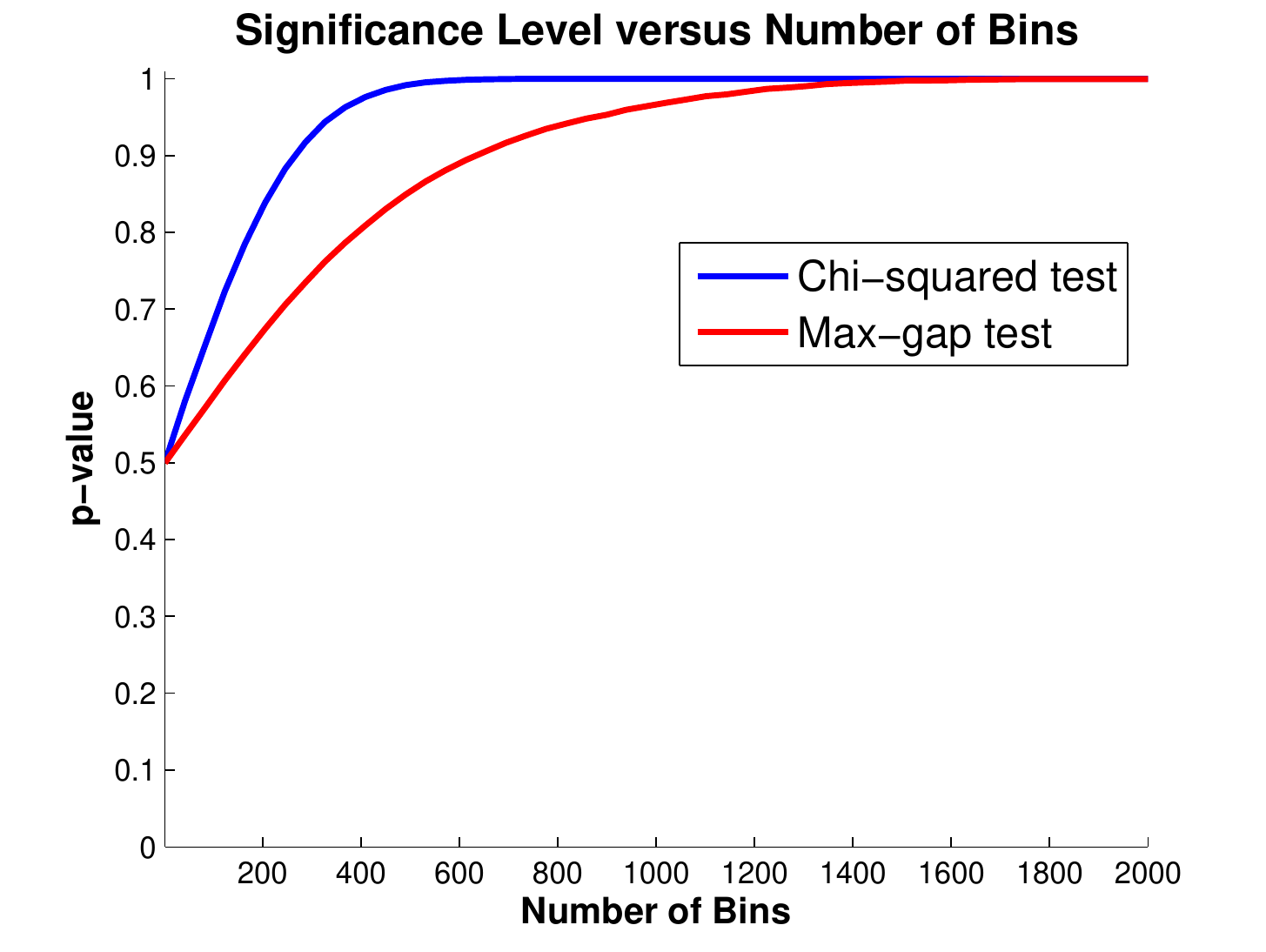}
\caption{p-value of the $\chi^2$ test and the max-gap test for high-regularity data. Regularity for a dataset of size $N$ is parameterized by a number of bins $k$ with $N/k$ uniform samples within each of $k$ equal-width bins. (Thus $k=1$ generates a uniform distribution and increasing $k$ approaches regular spacing.)  The max-gap test again demonstrates greater sensitivity.}
\label{fig:high-regularity}
\end{figure}
 
In this section, we compare the max-gap test versus the most well-known and commonly used $\chi^2$ test. We conducted four experiments involving datasets of $N=10,000$ samples, with the result for each experiment obtained as an average of one million independent tests. Sensitivity is assessed by comparing the respective $p$-values for the one-sided forms of the two tests, where smaller values indicate greater sensitivity. The first experiment was performed using a dataset of samples from a true uniform distribution. As expected, the dataset passed both tests for uniformity with $p = 0.5$.
 
The second experiment examined sensitivity to the difference between a uniform distribution and a normal distribution with standard deviation $\sigma$ sampled within a fixed interval $(0, 1)$. The distinctive shape of the normal distribution is realized within the interval when $\sigma$ is small but flattens with increasing values and approaches uniformity. Both tests are equally sensitive for small $\sigma$, and both approach $p=0.5$ for large $\sigma$, but the $\chi^2$ test exhibits higher sensitivity for intermediate values (see Fig. \ref{fig:variance}). The latter is not surprising because the $\chi^2$ test is ideally sensitive to deviations from normality.
 
The third experiment examined sensitivity of the two tests to a uniform distribution with a narrow-band exclusion (Fig. \ref{fig:missing-values}). This of course is a problem for which the max-gap test is ideally suited, and (\ref{eqn:expected-max-gap}) reveals that superior sensitivity. What is possibly most interesting about the results is that the $\chi^2$ test provides only modest sensitivity even as the exclusion width approaches one percent of the distribution window.
 
The fourth experiment is the most relevant for data integrity applictions. It examined sensitivity to regularity in sample spacing. Anomalous distribution regularity is a common characteristic of human-altered data because people typically underestimate the degree of natural ``clustering'' that is present in data sampled from a truly uniform distribution. As a consequence, human-created or human-altered data tends to have higher regularity, i.e., tends to be ``more evenly distributed'', than what is expected for uniformly-distributed data. More generally, high-regularity deviations from uniformity can arise from the unanticipated influence of a structured or non-random process, e.g., frequency-combing effects from a physical sensor or simulation artifacts resulting from a low-quality pseudorandom number generator.
 
A regularity parameter $1\leq k \leq N$ was used for this experiment by uniformly distributing $N/k$ samples within each of $k$ equal-width subdivisions of the distribution interval. Thus $k=1$ represents a uniform sampling over the entire interval and produces a uniform distribution; and as $k$ increases to $N$ the spacing between samples becomes increasingly regular. Although uniform and high-regularity distributions are difficult for humans to distinguish visually, Fig. \ref{fig:high-regularity} shows that the max-gap test provides significantly higher sensitivity than $\chi^2$ to subtle regularity deviations from uniformity.
 
\section{Min-Gap Test}
 
The one-sided variants of the max-gap and $\chi^2$ tests were used because they provide a practical balance between high sensitivity and low false alarm rates, but the one-sided or two-sided of either test may provide the optimal trade-off for the needs of a particular given application. In some applications the optimal trade-off might be obtained from a \emph{min-gap}, $S_{(1)}$, test. The min-gap approximated distribution is given by \cite{Holst1980}
\begin{equation}
P \left( S_{(1)} \leq x \right) \overset{N \rightarrow \infty}{=} e^{-e^{\ln N - Nx}} \sum\limits_{\nu = 0}^{N - 1} \frac{\left( e^{\ln N - Nx} \right)^\nu}{\nu!},
\label{eqn:min-gap-distribution}
\end{equation}
and its expected value is \cite{Holst1980}
\begin{equation}
E \left( S_{(1)} \right) \overset{N \rightarrow \infty}{=} \frac{\gamma + \ln N + \sum\limits_{i = 1}^{N - 1} \frac{1}{i}}{N}.
\label{eqn:expected-min-gap}
\end{equation}
 
A min-gap test can be defined and performed analogously to the max-gap test and would be ideally suited for detecting spuriously-replicated data items. However, simpler non-statistical methods can be applied to detect replicated data, so the potantial applications of the min-gap test may be somewhat more limited than the max-gap test.
 
\section{Computational Considerations}
 
In terms of computational complexity, both $\chi^2$ and max-gap tests can be evaluated in optimal $O(N)$ time and $O(N)$ space. This complexity is achieved for max-gap by use of the Gonzalez algorithm \cite{Gonzalez1975,Gonzalez1985} to determine the max-gap in linear time without sorting. The Gonzalez algorithm performs a special binning which guarantees by the pigeonhole principle that the max-gap data items will be found as the maximum and minimum values, respectively, in consecutive non-empty bins. This algorithm allows the max-gap test to be evaluated in optimal $O(N)$ time and space, i.e., the same as $\chi^2$, and is as efficiently parallelizable\footnote{The max-gap and $\chi^2$ tests are both highly amenable to parallelization with $O(N/P)$ time complexity on $P$ processors.} as the $\chi^2$ test.
 
The min-gap pair needed to implement a min-gap test can be identified in optimal expected $O(N)$ time and space using Rabin's randomized closest-pair algorithm\cite{Golin1995,Lipton2010}. Unlike the Gonzalez algorithm for max-gap, Rabin's algorithm generalizes efficiently to higher dimensions.
 
\section{Discussion and Future Work}
We have defined and developed a max-gap test for distinguishing deviations from uniformity in a 1D dataset of size $N$. By using Gonzalez's algorithm we have shown that this test can be performed with commensurate efficiency, both serial and in parallel, with the conventional $\chi^2$ test. Our experiments demonstrate that the max-gap test provides improved sensitivity in two particular applications of relevance to data integrity verification. More generally, the proposed max-gap and min-gap tests are of potential value as alternatives or to complement the use of $\chi^2$ for distribution testing and discrimination.
 
There are many statistical tests for equality of distributions beyond the $\chi^2$ test such as the Kolmogorov-Smirnov test \cite{Birnbaum1951,Conover1999,Marsaglia2003} and the Cramer-von Mises test \cite{Cramer1928}. Of course there can be no test that is uniformly superior to all others for all possible distributions, but it appears that most of the standard tests examined in the literature would be challenged similarly to the $\chi^2$ test to distinguish uniform from regularly distributed data.

Potential future work could consider tests which jointly combine gap and $\chi^2$ statistics into a more sophisticated single test \cite{Maynes2009} which allows greater flexibility to optimize the sensitivity and false alarm trade-off for problems of high practical interest, e.g., big data analytics and integrity verification. On the algorithmic side, we have pointed out that the Gonzalez algorithm does not generalize to higher dimensions; however, relatively efficient subquadratic algorithms do exist for solving the largest empty circle and largest empty rectangle problems in two dimensions \cite{Chazelle1986,Naamad1984}. Tests on $2d$ distributions could also potentially exploit information about the largest empty region of a Voronoi decomposition or the distribution of nearest-neighbor distances from a
Delaunay triangulation. In $d>2$ dimensions it may be possible to devise gap-related statistical tests based on results from efficient algorithms for identifying approximations to the largest empty $d$-sphere or $d$-rectangle, but this is purely speculative. In higher dimensions it may be better to abandon gap-type statistics and focus on statistics gleaned from efficiently-computable $k$-d and orthant (quad, octant, etc.) tree decompositions of point sets.
 
If computational efficiency is less of a concern, a perhaps more fruitful direction for highly-sensitive distribution testing in high dimensions is to examine the length of the Euclidean minimum spanning tree (EMST) for a dataset. The expected length of the EMST of uniformly-distributed points can be determined using analysis similar to what has been described in this paper for estimating the expected values for the max and min gaps in 1D, and we conjecture that EMST length is likely to be more sensitive to many practically important types of deviations from uniformity than the conventional $\chi^2$ test. Such an EMST test would be computationally expensive (though subquadratic), but this cost could be justified in applications for which subtle deviations are critically important, e.g., high-fidelity physics simulations.
 
\bibliographystyle{IEEETran}
\bibliography{references}
 
\end{document}